\documentclass[12pt]{article}

\usepackage{amsmath,amsfonts,amssymb,latexsym,graphicx,hyphenat}

\usepackage{caption}
\usepackage{subcaption}
\usepackage{mathtools}
\usepackage[T1]{fontenc}
\usepackage[utf8]{inputenc}
\usepackage{microtype}

\newtheorem{theorem}{Theorem}[section]

\newtheorem{remark}{Remark}[section]

\numberwithin{equation}{section}
\def\be{\begin{equation}}
\def\ee{\end{equation}}
\def\bq{\begin{eqnarray}}
\def\eq{\end{eqnarray}}
\def\beq{\begin{eqnarray}}
\def\eeq{\end{eqnarray}}

\def\a{\alpha}

\PassOptionsToPackage{dvipsnames}{xcolor}
\usepackage{xcolor}
\usepackage{silence}
\usepackage{tikz}
\usepackage{tikz-3dplot}
\usetikzlibrary{arrows.meta,calc,decorations.markings}
\begin{document}
\title{\textsc{Mode interactions in scalar field cosmology}}
\author{\Large{\textsc{Spiros Cotsakis$^{1,2}$\thanks{\texttt{skot@aegean.gr}} and Ignatios Antoniadis$^{3,4}$}}\thanks{\texttt{antoniad@lpthe.jussieu.fr}} \thanks{On leave from LPTHE, Sorbonne Universit\'{e}, CNRS, 4 Place Jussieu, 75005  Paris, France}\\
$^{1}$Clare Hall, University of Cambridge, \\
Herschel Road, Cambridge CB3 9AL, United Kingdom\\
$^{2}$Institute of Gravitation and Cosmology,  RUDN University\\
ul. Miklukho-Maklaya 6, Moscow 117198, Russia\\
$^{3}$Institute for Advanced Study\\ 1 Einstein Drive, Princeton, New Jersey 08540,  USA\\
$^{4}$High Energy Physics Research Unit, Faculty of Science\\ Chulalongkorn University\\ Phayathai Road, Pathumwan, Bangkok 10330, Thailand
}
\date{January 2026}
\maketitle
\newpage
\begin{abstract}
\noindent We study the dynamics of spatially homogeneous Friedmann--Robertson--Walker
universes filled with a massive scalar field in a neighbourhood of the massless
transition $s=1$. At this point the Einstein--scalar system exhibits a
codimension--two Hopf--steady--state organising centre whose versal unfolding
describes all small deformations of the quadratic model. After reduction to the
centre manifold, the dynamics is governed by two slow geometric modes $(r,z)$:
the Hopf amplitude $r$, measuring the kinetic departure from de Sitter, and the
slowly drifting Hubble mode $z$. We show that the standard slow--roll
parameters follow directly from these unfolding variables,
$\epsilon\sim\tfrac32 r^{2}$ and $\eta\sim z$, so that the spectral tilt,
tensor--to--scalar ratio, and scalar amplitude arise as universal functions of
$(r,z)$, independently of the choice of potential. The two unfolding parameters $(\mu_{1},\mu_{2})$ classify all perturbations of
the quadratic model and can be interpreted physically as controlling the tilt
and curvature deformations of generic polynomial inflationary potentials.
 Thus the near scale--invariance of primordial
perturbations emerges as a structural property of the unfolding of the
organising centre, providing a potential--independent mechanism for an early
phase of accelerated expansion. We discuss the implications of this geometric
framework for the interpretation and classification of inflationary models.
\end{abstract}
\newpage
\tableofcontents
\newpage

\section{Introduction}\label{sec:intro}
Scalar fields play a central role in early--universe cosmology,
both as effective
matter sources within general relativity and as carriers of dynamical degrees of
freedom arising from particle physics or modified gravity \cite{Guth1981,Linde1983,MukhanovBook} (see also \cite{SHS2014,ChervonBook} for complementary geometric/dynamical-systems perspectives on scalar field cosmology).  In the
Friedmann--Robertson--Walker (FRW) setting, the Einstein--scalar equations form a
finite-dimensional dynamical system whose qualitative properties can be studied
using the tools of bifurcation and singularity theory.  This viewpoint is part
of a broader programme in which geometrical transitions in GR are interpreted as
bifurcations of the underlying field equations
\cite{1,2,3,4,5}.

In this paper we revisit the simplest scalar--field cosmology, the FRW universe
with a quadratic potential, focusing on the behaviour near the
\emph{massless transition}. Although the original chaotic inflation model is
disfavoured by current observational constraints \cite{bicep}, various polynomial
extensions of it (for example $V\sim\tfrac12 m^2\phi^2+\lambda\phi^n$,
$n\in\mathbb{Z}_{>0}$) remain compatible with the cosmic microwave background
data (see also \cite{klw14,kl25} for other two-parameter extensions). Our goal
is not to advocate a particular potential; rather, we show that the
\emph{universal deformation} structure of the Einstein--scalar dynamics near a
nondegenerate vacuum already contains the geometric origin of inflationary
observables. Pure exponential potentials, which have no stationary point and
only scaling solutions, lie in a different universality class and are therefore
not covered by the present analysis.

We work with an $m$--independent scaling of the Einstein--scalar FRW equations,
introducing the structural parameter $s=m^{2}$ as a distinguished control
parameter.  The resulting system (Eqs.~(2.1)--(2.4)) is smooth in $s$ and admits
a codimension--two organising centre at $s=1$ (corresponding to the massless
threshold in the original variables).  At this point one real and two imaginary
eigenvalues become simultaneously neutral, producing a Hopf--steady--state mode
interaction.  The versal unfolding of this degeneracy yields two canonical
families (Cases~I and II), organised by saddle--node, pitchfork and Hopf
bifurcations, and in the cubic family gives rise to invariant tori describing
persistent coupled oscillations of the scalar and geometric modes.

The key result of the paper is that the slow geometric variables of the versal
unfolding provide a direct route to inflationary observables.  On the centre
manifold the dynamics admits a natural fast--slow decomposition: a fast Hopf
phase $\theta$ and two slow variables $(r,z)$, where the Hopf amplitude $r$
measures the kinetic departure from de Sitter while the slow Hubble mode $z$
governs the geometric drift.  We show in
Sect.~\ref{subsec:slowroll} that the Hubble slow--roll parameters
satisfy $\epsilon\sim\tfrac32 r^{2}$ and $\eta\sim z$, implying that the
spectral tilt, tensor amplitude and scalar power are universal functions of the
unfolding coordinates $(r,z)$ and of the deformation parameters
$(\mu_{1},\mu_{2})$.  For any analytic potential with a nondegenerate vacuum
the physical parameters of the model map to $(\mu_{1},\mu_{2})$; familiar
deformations that control tilt and plateau behaviour are thus realised as
coordinates on this canonical unfolding, rather than being tied to any
particular choice of potential.

The local normal form, its versal unfolding, and the associated bifurcation diagrams (together with the relations $\epsilon\sim\tfrac32 r^{2}$ and $\eta\sim z$ on the centre manifold) are universal features of the organising centre. By contrast, the map from a given analytic potential $V(\phi)$ to the unfolding parameters $(\mu_{1},\mu_{2})$, the trajectory followed in the $(\mu_{1},\mu_{2})$--plane, and the global exit from the unfolding neighbourhood (e.g. reheating and total e--folds) are model-dependent.

The paper is organised as follows. In Sect.~2 we formulate the
Einstein--scalar FRW system in $m$--independent variables and identify the
organising centre at $s=1$.  Sect.~3 constructs the Hopf--steady--state normal
form and its versal unfoldings, and derives the bifurcation diagrams for Cases~I
and II.  Sect.~4 interprets these structures cosmologically, including the
appearance of invariant tori and the universal relations for inflationary
observables.  We conclude in Sect.~5 with a discussion of the geometric origin
of inflation and the role of versal deformations in classifying inflationary
models.

\textbf{What is known about this problem.}
We consider an FRW universe with curvature index \(k=0,\pm 1\), sourced by a scalar field \(\phi\) with
the FRW line element (signature $(-,+,+,+)$)
\be
ds^{2}=-dt^{2}+a(t)^{2}\left(\frac{dr^{2}}{1-k r^{2}}+r^{2}d\Omega_{2}^{2}\right),
\ee
where $d\Omega_{2}^{2}=d\vartheta^{2}+\sin^{2}\vartheta\, d\varphi^{2}$, and with
energy density and pressure
\be
\rho = \tfrac12 \dot{\phi}^{2} + V(\phi),\qquad
p = \tfrac12 \dot{\phi}^{2} - V(\phi),
\ee
and with the classical massive potential
\be
V(\phi)=\tfrac12 m^{2}\phi^{2}.
\ee
Note that adding a cosmological constant as a second distinguished parameter (in addition to $m^2$) produces a genuinely more intricate structural perturbation problem, which we do not address here (cf. \cite{2} for a complete analysis in the simplest FRW–fluid setting). Of course,  exact solutions are also available for the shifted quadratic potential $V(\phi)=\tfrac12 m^{2}\phi^{2}+\Lambda$; see, e.g., \cite{ChervonBook} and references therein.

The Einstein–scalar system is then governed by the evolution equations (a dot denotes differentiation
with respect to proper time and throughout we work in reduced Planck units $M_{\rm Pl}=1$),
\begin{align}
\ddot{\phi} + 3H\dot{\phi} + m^{2}\phi &= 0, \label{eq:phi_evol}\\[2pt]
\dot{H} + H^{2} &= \tfrac16 m^{2}\phi^{2} - \tfrac13 \dot{\phi}^{2}, \label{eq:Hdot}
\end{align}
together with the Friedmann constraint
\begin{equation}\label{eq:Friedmann}
H^{2} + \frac{k}{a^{2}} = \tfrac16\!\left(\dot{\phi}^{2} + m^{2}\phi^{2}\right).
\end{equation}

Classically, the transformation introduced in Refs.~\cite{6,7,8,9} rescales the variables using the
mass parameter \(m\),
\be\label{trans}
x=\frac{\phi}{\sqrt6},\qquad
y=\frac{\dot{\phi}}{\sqrt{6}\,m},\qquad
z=\frac{H}{m},\qquad
\tau=m t,
\ee
thereby removing \(m\) from the evolution equations while retaining it only in the constraint.
In these variables the evolution system becomes
\begin{equation}\label{eq:old_system}
x' = y,\qquad
y' = -x - 3yz,\qquad
z' = x^{2} - 2y^{2} - z^{2},
\end{equation}
with constraint
\begin{equation}\label{eq:old_constraint}
x^{2} + y^{2} - z^{2} = \frac{k}{m^{2}a^{2}}.
\end{equation}

This formulation underlies the classical analyses of Belinski–Grishchuk–Khalatnikov and
Gibbons–Hawking–Stewart.
It partitions the phase space into trajectories lying on the cone \(z^{2}=x^{2}+y^{2}\) (flat models),
in its interior (open models), and in its exterior (closed models), reflecting the geometry of the
constraint~\eqref{eq:old_constraint}.
The dynamics is essentially hyperbolic: linearisation yields a stable focus at the origin for \(k=0\),
corresponding to the late–time approach to an inflationary epoch,
together with four hyperbolic equilibria at infinity representing early–time behaviour, including
the big bang.

A characteristic limitation of this classical treatment is its reliance on hyperbolicity and linear
stability.
Eternally oscillating solutions exist but require finely tuned initial conditions.
Inflation arises generically in this framework, but the picture is dominated by the stable–focus
structure of the origin in the flat case and the global geometry encoded in the Friedmann constraint.

Recent observational constraints disfavour the pure quadratic potential, but polynomial
extensions (notably two–parameter families) have been shown to remain compatible with all
current CMB data \cite{bicep,klw14,kl25}.
The present work re-examines the foundational massive model from the perspective of \emph{versal
unfoldings} and \emph{mode interactions}, by replacing the above transformation with a new scaling
independent of \(m\).
This introduces the single distinguished parameter
\(
s = m^{2},
\)
and leads to a smooth one–parameter family of dynamical systems through which the transition
\(s>0 \rightarrow s<0\) (massive $\leftrightarrow$ tachyonic) can be analysed within a unified bifurcation
framework.
This new formulation is developed in Section~3.

\section{Summary of the main results}

\paragraph{Dimensionless form and control parameter.}
To obtain a unified formulation valid for both massive (\(m^{2}>0\)) and tachyonic (\(m^{2}<0\)) scalar fields,
we fix the scaling independently of \(m\) and introduce the single distinguished parameter
\be
s := m^{2}\in\mathbb{R},
\ee
so that \(s>0\) corresponds to a convex potential, \(s=0\) to the degenerate case, and \(s<0\) to a
tachyonic (negative mass-squared) or hill–top potential.  Throughout the kinetic term remains canonical (no ghost instability is implied).
We then define the dimensionless variables
\begin{equation}\label{eq:dimless_vars}
x=\frac{\phi}{\sqrt{6}},\qquad
y=\frac{\dot{\phi}}{\sqrt{6}},\qquad
z=H,\qquad
\tau=t,
\end{equation}
so that a prime denotes differentiation with respect to~\(\tau\).
In these variables, the Einstein–scalar field equations \eqref{eq:phi_evol}–\eqref{eq:Hdot} reduce to the single compact system
\begin{equation}\label{eq:system_s}
x' = y,\quad
y' = -s\,x - 3yz,\quad
z' = s x^{2} - 2 y^{2} - z^{2},
\end{equation}
together with the constraint \eqref{eq:Friedmann}, now written as
\begin{equation}\label{eq:constraint_s}
s x^{2} + y^{2} - z^{2} = \frac{k}{a^{2}}.
\end{equation}
This form is smooth in the single real parameter \(s\) and allows a unified bifurcation analysis
around the degenerate value \(s=0\), which marks the transition from oscillatory (\(s>0\)) to unstable (\(s<0\)) scalar–field dynamics.
\begin{remark}[On Eq.~(2.2) and the role of $s$]Equation~\eqref{eq:dimless_vars} only defines the dimensionless \emph{state} variables $(x,y,z)$ and the time variable $\tau$. The distinguished physical parameter $s=m^{2}$ is \emph{not} absorbed into these variables (in contrast to the classical rescaling \eqref{trans}), but appears explicitly in the evolution system \eqref{eq:system_s} and the constraint \eqref{eq:constraint_s}. \end{remark}
\paragraph{Linear structure and degeneracy.}
For the specific value \(s=1\), the linear part of system~\eqref{eq:system_s} at the origin is
equivalent to the matrix
\be
\begin{pmatrix}
0 & -\omega & 0\\[1pt]
\omega & 0 & 0\\[1pt]
0 & 0 & 0
\end{pmatrix},
\ee
so that the equilibrium at the origin has one zero and two purely imaginary eigenvalues \(\pm i\omega\).
This simultaneous presence of a zero and a pair of imaginary eigenvalues produces a
\emph{Hopf–steady–state interaction}, the fundamental nonlinear mechanism underlying the mode interactions analyzed in this paper.
The eigenvalue crossing at \(s=0\) triggers a qualitative change in stability and produces a three–dimensional centre manifold on which all local dynamics unfold.

Below we use the term \emph{organising centre} in the standard bifurcation-theory sense: the normal form reduction of the originally given system of dynamical equations that possesses a \emph{nonhyperbolic equilibrium at a distinguished parameter value} whose (versal) unfolding organises the nearby local phase portraits. In the present problem it is the system \eqref{eq:organizingcenter} with equilibrium $(x,y,z)=(0,0,0)$ at $s=1$ (equivalently at the shifted parameter value $\mu_{0}=s-1=0$, where the linear oscillatory frequency is normalised to unity), with eigenvalues $0$ and $\pm i$, giving a Hopf--steady--state (zero--Hopf) interaction.


With our $m$--independent scaling, $s=m^{2}$ is a genuine control parameter, so the physical massless limit corresponds to $s=0$. At the organising centre relevant for the Hopf--steady--state interaction  at $s=1$, where the linear oscillatory frequency is normalised to unity,  we  use the detuning $\mu_{0}=s-1$ in the normal form calculation (cf. Sect.~3). The terminology \emph{massless threshold} for $s=1$ refers to the following elementary observation: starting from \eqref{eq:system_s}, the massless equations are obtained either by setting $s=0$ (i.e.\ $m=0$), or, equivalently, by taking $s=1$ and setting $x=0$ in the $s$--dependent terms of \eqref{eq:system_s}. This equivalence is already implicit in the classical Gibbons--Hawking--Stewart formulation  \cite{8} and explains why we refer to $s=1$ as the massless threshold in the organising centre normalisation.

\paragraph{Versal unfolding and reduced systems.}
Carrying out the standard normal–form analysis (following Refs.~\cite{13,14,15,16,17} and using the Fredholm alternative as in~\cite{2}), the reduced system on the centre manifold
is conveniently expressed in cylindrical coordinates \((r,\theta,z)\) and takes the versal form
\begin{equation}
\dot{z} = \mu_{1} + z^{2} + d\, r^{2}, \qquad
\dot{r} = \mu_{2} r + \frac{3}{2}\, r z, \qquad
\dot{\theta} = \omega + c_{1} z,
\end{equation}
where \(d=\pm 1\) distinguishes the two main cases: \(d=+1\) for \(s>2\) (quadratic reduction sufficient),
and \(d=-1\) for \(s<2\) (cubic terms required).
The two unfolding parameters \(\mu_{1},\mu_{2}\) encode the leading perturbations of the degenerate linear part. Here $(r,\theta,z)$ are \emph{state-space} coordinates (Hopf amplitude,
phase and slow background variable) for the equilibrium at the origin,
whereas $\mu_{1},\mu_{2}$ denote \emph{external} physical parameters
(e.g. Taylor coefficients of $V(\phi)$). To remain within the validity of the centre-manifold reduction, one must restrict to a neighbourhood
\be
|x|,|y|,|z|\leq \varepsilon,\qquad |s|\leq c\,\varepsilon,
\ee
with $0<\varepsilon\ll1$ and $c=O(1)$. Here $s$ has been shifted so that the
organising-centre value of the original parameter (the massless threshold
$s=1$) corresponds to $s=0$; in these rescaled variables the limit $s\to 0$
captures the degenerate Hopf--steady--state case.
This ensures that the $O(\varepsilon^{2})$ terms in the vector field are of comparable magnitude
and that the degenerate limit $s\to0$ is captured correctly.
Within this neighbourhood, the versal unfoldings \eqref{eq:CaseI}–\eqref{eq:CaseII} provide a
complete description of the local dynamics and its physical consequences.
Eliminating the rotational variable \(\theta\) yields two effective two–dimensional versal families:
\begin{align}
\text{Case I ($d=+1$):}\qquad
&\dot{z} = \mu_{1} + z^{2} + r^{2},\qquad
\dot{r} = \mu_{2} r + \frac{3}{2} r z; \label{eq:caseI}\\[4pt]
\text{Case II ($d=-1$):}\qquad
&\dot{z} = \mu_{1} + z^{2} - r^{2},\qquad
\dot{r} = \mu_{2} r + \frac{3}{2} r z + \ell z^{3}, \label{eq:caseII}
\end{align}
with \(\ell\) a constant determined by the third–order terms.
Setting \(\mu_{1}=\mu_{2}=0\) produces the organizing centres that generate the full stratified bifurcation structure shown in Fig.~\ref{fig:bifn}.

\paragraph{Bifurcation diagrams and invariant sets.}
The resulting bifurcation diagrams are shown in Fig.~\ref{fig:bifn} and their physical interpretation in terms of slow–roll parameters is discussed in Sect.~\ref{sec:phys}\ref{subsec:slowroll}. They consist of seven strata in Case~I and eleven in Case~II
(see Fig.~\ref{fig:bifn}), corresponding to topologically distinct phase portraits organised by the
versal unfolding.
A key feature is the emergence and persistence of invariant tori in Case~II, a direct nonlinear
manifestation of Hopf–steady–state mode interaction in the full three–dimensional system~\eqref{eq:system_s}.
These tori arise from the limit cycles along the \(\alpha\)-stratum of the bifurcation diagram
and signal the presence of mixed geometry–matter oscillatory dynamics not detected by
classical hyperbolic or linearised analyses.

\paragraph{How to read the bifurcation diagrams.}
For physical interpretation it is useful to state how slow--roll (SR), ultra
slow--roll (USR) and oscillatory regimes appear in the diagrams of
Fig.~\ref{fig:bifn}.  On the centre manifold, the asymptotically stable invariant
set in a given stratum determines the regime: a single hyperbolic stable
equilibrium corresponds to a slow--roll attractor, while a stable limit cycle
(on the $\alpha$--stratum of Case~II) yields a persistently oscillatory state
and, in the full three--dimensional system, an invariant torus.  In terms of
the unfolding coordinates we have $\epsilon\sim\tfrac32 r^{2}$ and
$\eta\sim z$, so hyperbolic strata with a unique stable focus at small $(r,z)$
(such as the $\chi$--region in Case~II) reproduce the standard SR fixed point
in the $(\phi,\dot\phi)$--plane, whereas the $\gamma$--stratum (containing the
origin) together with its neighbouring $\eta$--stratum around
$(\mu_{1},\mu_{2})=(0,0)$ represent the nonhyperbolic USR organising centre
with $\epsilon\ll1$ and $|\eta|=O(1)$.

Along the $E\rho$ semi--axis the system admits a nonhyperbolic continuum of
periodic orbits (like the undamped harmonic oscillator), a finely tuned
situation destroyed by generic perturbations.  On the $\alpha$--stratum this
degeneracy is resolved into a single hyperbolic stable limit cycle,
representing a robust oscillatory phase and, in the full system, a persistent
invariant torus.  Typical parameter paths then realise sequences such as
$\chi\to E\rho\to\alpha$, corresponding to
SR~$\to$~a mixed multi--equilibrium regime on the $E\rho$ semi--axis~$\to$~a
single--frequency oscillatory state generated by the limit cycle, or
$\eta\to\alpha$, corresponding to USR~$\to$~oscillatory dynamics.  In this way
the transitions SR~$\leftrightarrow$~USR~$\leftrightarrow$~oscillatory regimes
in scalar--field cosmology are seen to be organised by the stratified structure
of the versal unfolding, and reversing the motion of $(\mu_{1},\mu_{2})$ along
a path simply inverts the corresponding physical sequence (e.g.\
$\alpha\to E\rho\to\chi$ or $\alpha\to\eta$).

A further distinctive feature of Case~II is the presence of the blow-up curve
$\nu$.  As one traverses small loops in the unfolding plane around the
organising-centre point, the stable limit cycle on the $\alpha$--stratum grows
in amplitude and eventually reaches the boundary of the local region of validity
of the centre-manifold reduction along $\nu$, realising a cycle blow-up in the
sense of Ref.~\cite{15}.  This curve thus mediates the transition between
small-amplitude oscillations and large excursions in $(r,z)$ within the versal
unfolding.

\section{The versal family: Dynamics and bifurcations}

\subsection{The normal form and versal unfolding}

We now analyse the local dynamics of the system \eqref{eq:system_s} near the origin for the
distinguished parameter value \(s=1\).
As noted in Section~2, the equilibrium at \((x,y,z)=(0,0,0)\) has a degenerate linear part
consisting of one zero and two purely imaginary eigenvalues.
This configuration is neither hyperbolic nor reducible to a standard centre–saddle form; instead
it represents a \emph{Hopf–steady–state} mode interaction in which two oscillatory modes and one
slow mode become simultaneously neutral.
In this subsection we derive the associated normal form and its versal unfolding.

\paragraph{Rewriting the system.}
To place \eqref{eq:system_s} into the canonical form used in normal–form theory, it is convenient
to interchange \(x\leftrightarrow y\) and rewrite the oscillatory part as
\[
y'' + s\,y = 0,
\]
so that the linear oscillations take the standard form.
Introducing \(x=y'\), and denoting by \(f,g,h\) the nonlinearities, the system becomes
\be
x' = -s\,y + f(x,y,z), \qquad
y' = \phantom{-}x + g(x,y,z), \qquad
z' = h(x,y,z),
\ee
where, from \eqref{eq:system_s},
\begin{equation}\label{eq:fghexplicit}
f=-3xz,\qquad
g=0,\qquad
h=s\,y^{2} - 2x^{2} - z^{2},
\end{equation}
together with the constraint
\begin{equation}\label{eq:constraintNF}
x^{2} + s\,y^{2} - z^{2} = \frac{k}{a^{2}}.
\end{equation}

\paragraph{Reduction to the centre manifold.}
Following the standard procedure (cf.\ Refs. \cite{13}-\cite{17}), we shift the parameter to the origin
\(
\mu_{0}=s-1,
\)
identify a basis of generalized and adjoint eigenvectors of the linear part,
and apply the Fredholm alternative as in~\cite{2}.
This yields a three–dimensional centre manifold parametrised by
$(r,\theta,z)$ in cylindrical coordinates, capturing all small perturbations
of the degenerate equilibrium.

Truncating to second order, the reduced dynamics on the centre manifold takes the form
\be\label{eq:NFzrtheta}
z' = b_{1} r^{2} + b_{2} z^{2} + O(3),\qquad
r' = a_{1} r z + O(3),\qquad
\theta ' = 1 + c_{1} z + O(2),
\ee
where explicit calculation using~\cite{18} yields
\be
a_{1}=-\frac{3}{2},\qquad
b_{1}=-\frac{\mu_{0}+3}{2},\qquad
b_{2}=-1,\qquad
c_{1}=0.
\ee

\paragraph{Normalisation.}
A final rescaling (following the conventions in~\cite{14}) introduces the sign
\[
b=\frac{-b_{1}b_{2}}{|b_{1}b_{2}|}
=
\begin{cases}
+1,& s<2,\\[2pt]
-1,& s>2,
\end{cases}
\]
and rewrites the normal form compactly as
\begin{equation}\label{eq:organizingcenter}
z' = b r^{2} - z^{2},
\qquad
r' = -\frac{3}{2} r z.
\end{equation}
Introducing \(T=-\tau\) (so that $\dot{}$ now denotes differentiation with respect to $T$)
places \eqref{eq:organizingcenter} into the standard form used for the two versal cases below.

\subsection{Versal unfoldings and bifurcation structure}

Adding the two unfolding parameters \(\mu_{1},\mu_{2}\) and including the cubic term
needed when \(b=+1\) yields the versal families:

\paragraph{Case I (\(s>2\), quadratic truncation sufficient).}
\begin{equation}\label{eq:CaseI}
\dot{z} = \mu_{1} + z^{2} + r^{2},
\qquad
\dot{r} = \mu_{2} r + \frac{3}{2}\,r z.
\end{equation}

\paragraph{Case II (\(s<2\), cubic term required).}
\begin{equation}\label{eq:CaseII}
\dot{z} = \mu_{1} + z^{2} - r^{2},
\qquad
\dot{r} = \mu_{2} r + \frac{3}{2}\,r z + \ell z^{3},
\end{equation}
where \(\ell\) is a nonzero constant determined by the coefficient of the cubic term in the
centre–manifold reduction; its explicit value is not needed in what follows.

Both \eqref{eq:CaseI} and \eqref{eq:CaseII} are \emph{versal unfoldings} of the degenerate
organizing centre \eqref{eq:organizingcenter}: all sufficiently small perturbations of
\eqref{eq:system_s} near \(s=1\) are locally equivalent to one of these two families.

\subsection{Structure of the bifurcation diagrams}

\begin{figure}[t]
\centering
\includegraphics[width=0.45\textwidth]{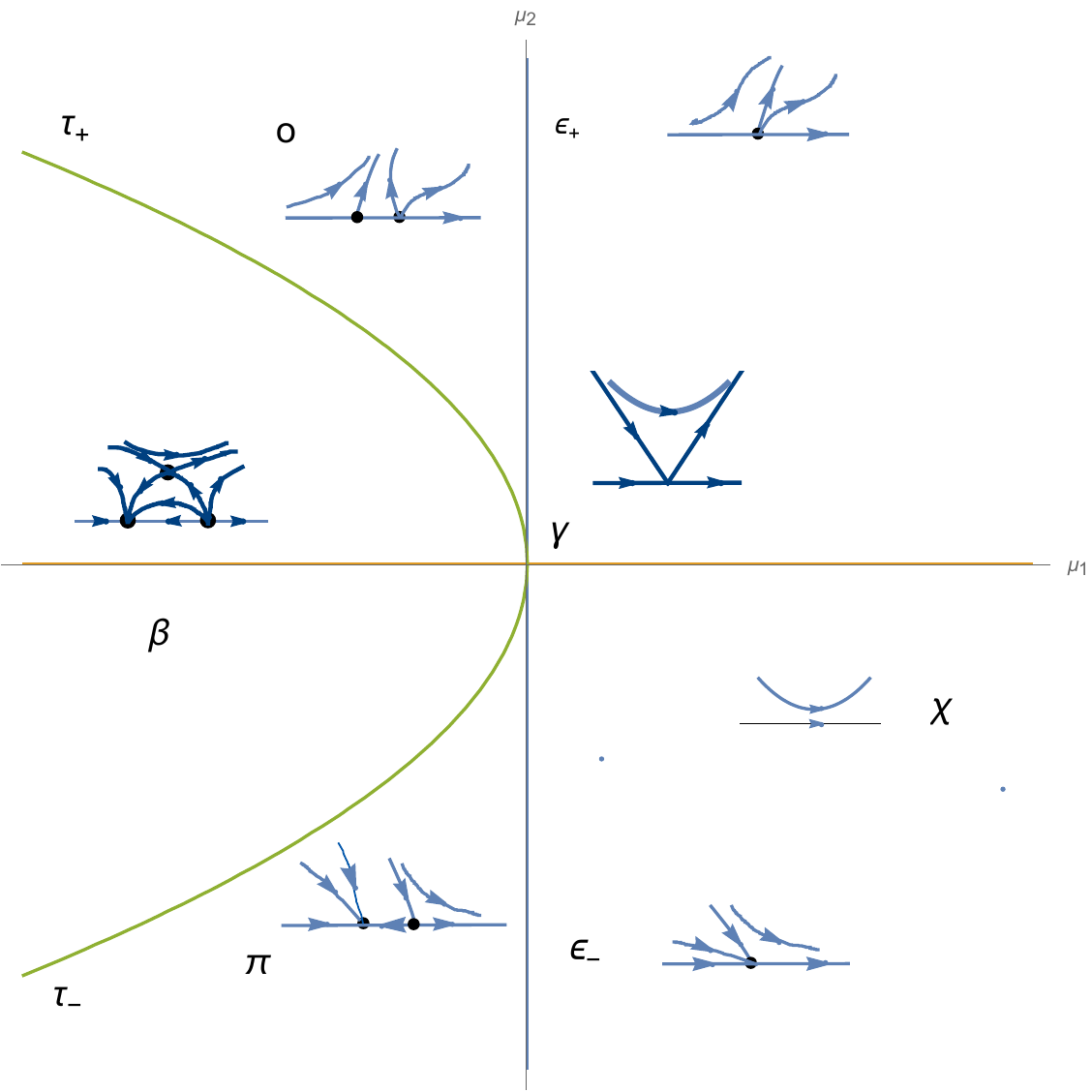}\qquad
\includegraphics[width=0.45\textwidth]{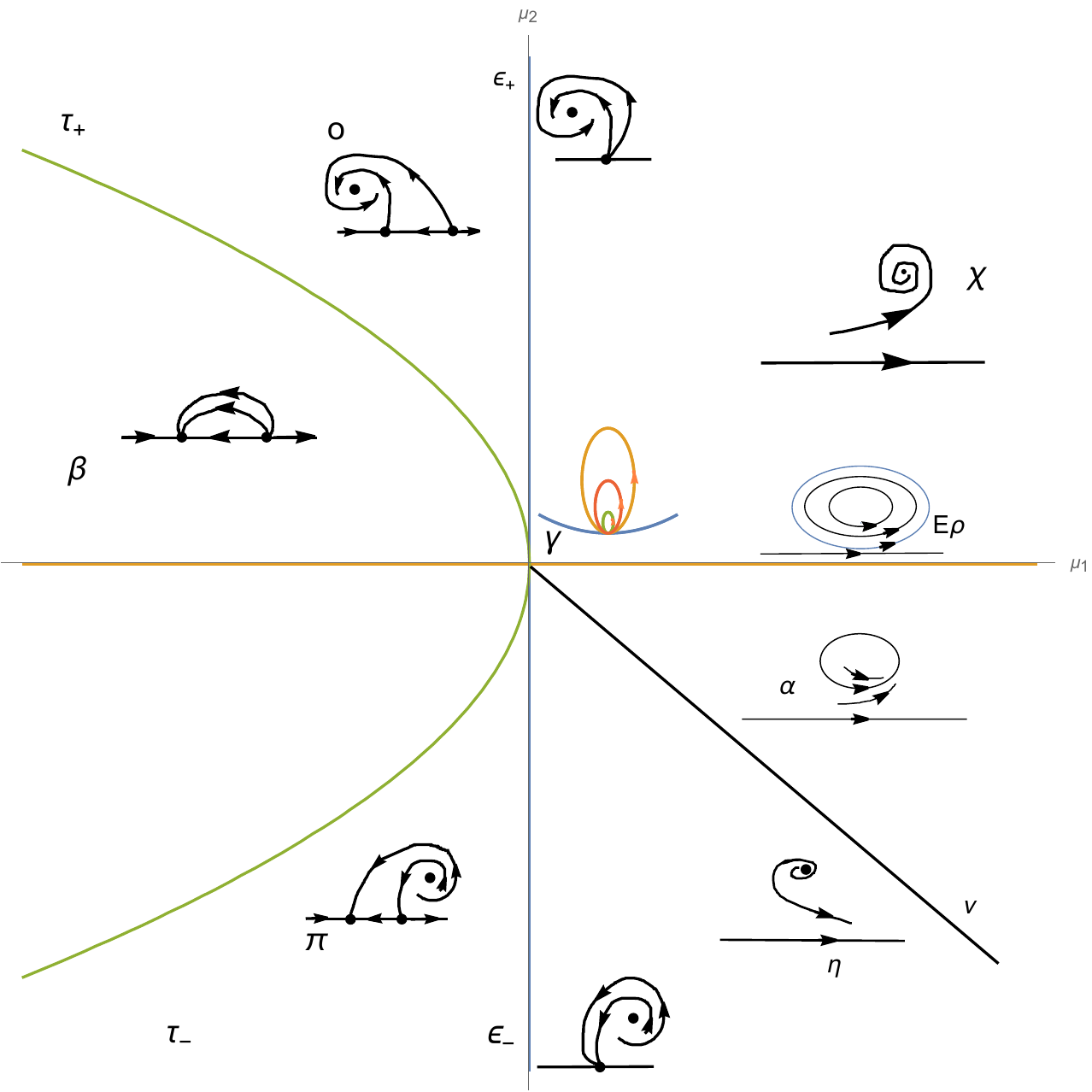}
\caption{%
Bifurcation diagrams for the versally unfolded FRW–scalar field system.
\textbf{Left:} The quadratic Case~I (\(s>2\)), exhibiting seven strata
\(\chi,\gamma,\varepsilon,o,\pi,\tau,\beta\).
\textbf{Right:} The cubic Case~II (\(s<2\)), with eleven strata
\(\chi,\gamma,E_{\rho},\alpha,\nu,\eta,\varepsilon,o,\pi,\tau,\beta\).
Each stratum corresponds to a region in the unfolding parameter plane \((\mu_{1},\mu_{2})\)
with a topologically distinct phase portrait.
The diagrams illustrate the organisation of saddle–node, pitchfork, and Hopf bifurcations
that arise from the Hopf–steady–state mode interaction of the FRW–scalar field system.}
\label{fig:bifn}
\end{figure}

The bifurcation diagrams associated with \eqref{eq:CaseI} and \eqref{eq:CaseII} are shown in
Fig.~\ref{fig:bifn}.
In Case~I there are seven strata, while in Case~II there are eleven; each stratum corresponds to a
region in the parameter plane \((\mu_{1},\mu_{2})\) where the phase portrait of the system is
topologically equivalent.

A useful set of first integrals of the centre-manifold dynamics at both
organising centres (i.e.\ for $\mu_{1}=\mu_{2}=0$) is given by
\begin{equation}\label{eq:firstintegrals}
I_{n}(z,r)=\frac{n}{2}\, r^{2/n}
\left(
z^{2} \pm r^{2}
\right)^{1-n},
\end{equation}
with \(n=\tfrac{3}{2}\) and the upper sign for Case~I, the lower for Case~II.
For these organising-centre systems the integrals provide a complete
description of the orbits in a neighbourhood of the equilibrium and recover, as
special cases, the stable–focus behaviour of the flat FRW model found in
Refs.~\cite{7,8}.

An important nonlinear feature of the Case~II unfolding is the phenomenon of
cycle blow-up in the sense of Kuznetsov~\cite{15}.  Along parameter
loops encircling the organising-centre point, a small limit cycle born in a
Hopf bifurcation on the $\alpha$--stratum can grow and reach the boundary of
any fixed neighbourhood of the origin before disappearing from the local
centre-manifold description along the curve $\nu$ (the analogue of Kuznetsov's
$J$-curve).  Cosmologically this means that arbitrarily small deformations of
the effective parameters $(\mu_{1},\mu_{2})$ at the organising centre
(corresponding to the massless transition) can drive the coupled
scalar--geometry oscillations from negligible amplitude to order-one excursions
in $(r,z)$, sharply amplifying the sensitivity of post-inflationary dynamics
to microphysical details of the potential.

\subsection{Fixed branches and bifurcations}
The two versal families admit up to three fixed branches in a neighbourhood of \((\mu_{1},\mu_{2})=(0,0)\)
that unfold the Milne equilibrium of the FRW system.
For both cases, the primary branches are
\begin{equation}\label{eq:E12}
E_{1,2}=(z^{(0)}_{1,2},0) = \left(\mp\sqrt{-\mu_{1}},\, 0\right),
\end{equation}
which exist for $\mu_{1}<0$.
A third branch appears in each case:
\paragraph{Case I.}
\begin{equation}\label{eq:ZI}
Z_{I} =
\left(
-\frac{\mu_{2}}{n},
\;
\sqrt{-\frac{\mu_{2}^{2}}{n^{2}} - \mu_{1}}
\right),
\qquad
\text{real when }
\mu_{1} < -\frac{\mu_{2}^{2}}{n^{2}}.
\end{equation}
\paragraph{Case II.}
\begin{equation}\label{eq:ZII}
Z_{II} =
\left(
-\frac{\mu_{2}}{n},
\;
\sqrt{\frac{\mu_{2}^{2}}{n^{2}} + \mu_{1}}
\right),
\qquad
\text{real when }
\mu_{2}^{2}/n^{2}+\mu_{1}>0.
\end{equation}
The stability and bifurcation properties of these branches follow directly from the linearisation and
have been extensively analysed in~\cite{4}.
We summarise the results here:
\begin{theorem}[Stability of the fixed branches]
The equilibria \(E_{1,2}, Z_{I}, Z_{II}\) exhibit the following behaviour:
\begin{enumerate}
\item
\(E_{1}\) is a saddle, sink, or bifurcates when
\(\mu_{2}-n\sqrt{-\mu_{1}}\) is \(>0\), \(<0\), or =0, respectively.
\item
\(E_{2}\) is a source, saddle, or bifurcates when
\(\mu_{2}+n\sqrt{-\mu_{1}}\) is \(>0\), \(<0\), or =0, respectively.
\item
\(Z_{I}\) is always a saddle (Case~I) and limits to \(E_{1}\) or \(E_{2}\) depending on whether the
\(\tau^{+}\) or \(\tau^{-}\) branch is followed.
\item
\(Z_{II}\) (Case~II) exhibits:
\begin{enumerate}
\item real eigenvalues: source for \(\mu_{2}<0\), sink for \(\mu_{2}>0\);
\item complex eigenvalues: unstable node if \(\mu_{2}<0\), stable node if \(\mu_{2}>0\);
\item purely imaginary eigenvalues at \(\mu_{2}=0,\,\mu_{1}>0\): a degenerate Hopf bifurcation producing an infinite family of closed orbits, stabilised by cubic terms.
\end{enumerate}
\end{enumerate}
\end{theorem}
\paragraph{Geometric interpretation.}
Crossing the various strata induces three characteristic bifurcations:
\begin{itemize}
\item \emph{Saddle--node (horizontal, $z$--direction)}: creation or annihilation of \(E_{1,2}\).
\item \emph{Pitchfork (vertical, $r$--direction)}: appearance of the branch \(Z_{I}\) in Case~I.
\item \emph{Hopf and torus formation (diagonal directions)}: in Case~II, the branch \(Z_{II}\)
undergoes a Hopf bifurcation, creating a limit cycle which lifts to an invariant torus of the
full three–dimensional system.
\end{itemize}
These bifurcations account for the geometry of the strata shown in Fig.~\ref{fig:bifn} and
underlie the coupled scalar–field and geometric oscillations generated by the Hopf–steady–state interaction, whose detailed behaviour is analysed below.

\section{Physical implications of the versal unfolding}\label{sec:phys}
The bifurcation diagrams obtained in the previous section summarise the
local phase–space organisation of the Einstein–scalar flow near the
Hopf–steady–state organising centre at $s=1$.
Each stratum of the versal unfolding corresponds to a distinct dynamical
regime, characterised by specific configurations of equilibria, periodic
orbits, and—in the cubic case—invariant tori.
These structures determine the qualitative behaviour of cosmological
solutions in a neighbourhood of the degeneracy, and their arrangement
in the unfolding plane provides a geometric framework for understanding
the various possible evolutionary paths of FRW scalar–field universes.

In this section we outline the physical interpretation of these regimes.
We begin by describing the dynamical significance of the fixed branches
and invariant sets, then discuss how the mode interaction constrains the
behaviour of the Hubble variable and the scalar field.
We finally show how the observable quantities $(n_{s},r_{s},A_{s})$
emerge directly from the unfolding variables $(z,r)$, providing a
geometric explanation for the robustness of inflationary behaviour.

\subsection{Interpretation of the strata}
\paragraph{Interpretation.}
The physical meaning of the dynamical structures in the versal unfolding is most
transparent when discussed separately for the two organising families.

\medskip
\noindent\textbf{Case I (quadratic family).}
In this regime the fixed branches $E_{1,2}$ and the parabolic curve $Z_{I}$
govern the evolution near the organising centre.
The dynamics is dominated by the competition between the slow geometric mode $z$
and the scalar oscillation amplitude $r$, with no higher–order geometric
feedback.
Trajectories typically approach one of the fixed branches, corresponding either
to slow contraction ($z<0$), slow expansion ($z>0$), or near–critical evolution
along the fold of the parabola.
The absence of cubic terms suppresses the formation of secondary oscillatory
structures: no invariant tori occur in this family.
Cosmologically, Case~I represents universes whose dynamics is controlled by a
single dominant mode (expanding or contracting) with weak scalar–field
interaction, producing monotonic or weakly damped behaviour of the Hubble rate.

\medskip
\noindent\textbf{Case II (cubic family).}
The cubic term in the $r$–equation induces a qualitatively richer set of
dynamical behaviours.
In addition to the fixed branches $E_{1,2}$, the cubic family contains the
$\mathit{cusp}$–shaped curve $Z_{II}$ and supports the emergence of invariant
tori through a secondary Hopf mechanism.
These tori correspond to persistent coupled oscillations of the geometric mode
$z$ and the scalar–field oscillation amplitude $r$, representing sustained
geometry–matter interactions.
Such mixed oscillatory states do not appear in Case~I.
Cosmologically, Case~II therefore describes universes in which the scalar and
geometric modes remain dynamically entangled for long periods, producing
quasi–periodic evolution of the Hubble parameter and the scalar energy density.
This richer oscillatory behaviour is a signature of the cubic unfolding and is
structurally stable within its domain.

\medskip
\noindent
In both cases the neighbourhood of $s=1$ forces the system into regimes with
small $r$ and small $z$, yielding the universal slow–roll expressions derived
below.
Thus the geometric origin of near–inflationary evolution holds across both
versal families, despite their very different phase–space structures.
It is important to note that the physically realised Einstein–scalar system lies
on the cubic side of the unfolding.  In a neighbourhood of $s=1$, the normal-form
coefficient determining the sign of the unfolding is negative, so that the
dynamics unfolds within the Case~II family.  This explains why invariant tori and
mixed scalar–geometric oscillations are generic in the physical problem.
Moreover, the massless case $s=0$ may be viewed as lying on a lower-dimensional
degenerate subset of this unfolding: when the $x$--dependence is suppressed at
$s=1$, the resulting flow reproduces the qualitative structure of the massless
system.  Thus the $s=0$ dynamics is naturally embedded in the geometry of the
organising centre at $s=1$.

\subsection{Implications for the full three-dimensional system}

Although the reductions \eqref{eq:CaseI}–\eqref{eq:CaseII} are two-dimensional, their bifurcation structures
lift to the full three-dimensional system \eqref{eq:NFzrtheta}, where the equilibria $E_{1,2}$ correspond to equilibrium \emph{points}, the nontrivial fixed branch $Z_{I}$ corresponds to a limit cycle of the same stability,  and the limit cycle in $(r,z)$ corresponds to an invariant torus.
In particular:

\begin{itemize}
\item The saddle-node creation of $E_{1,2}$ corresponds to changes in the qualitative behaviour of
$H$ and $\dot{\phi}$, determining whether the model is attracted to inflationary expansion or
diverges away from it.

\item The pitchfork birth of $Z_{I}$ marks a transition from geometry-dominated to matter-dominated
behaviour on the centre manifold, shifting the equilibrium among curvature, kinetic, and potential
terms.

\item The Hopf bifurcation of $Z_{II}$ produces a limit cycle in the reduced system, which lifts to an
\emph{invariant torus} in the full system, signalling the onset of sustained quasi-periodic
oscillations in $(\phi,\dot{\phi},H)$.
\end{itemize}

These effects are invisible to purely linearised analyses and arise only through a fully nonlinear
treatment of the degeneracy at $s=1$.

In the cubic unfolding (Case~II), the periodic orbit on the $(z,r)$–plane lifts to a
two–torus when rotated along the Hopf phase $\theta$.
This is the standard mechanism illustrated in various references (e.g., \cite{14,15}): a
periodic orbit in the reduced system $(z,r)$ generates an invariant torus
$S^{1}\times S^{1}$ in the full three–dimensional flow.

Cosmologically, such tori correspond to persistent quasi–periodic
oscillations of both the scalar–field kinetic energy and the Hubble
parameter.
The universe undergoes long–lived, non–chaotic oscillations in which the
scalar and geometric modes remain dynamically coupled, resulting in a
continuous, modulated exchange of energy between matter and geometry.
These quasi–periodic states arise purely from the geometric structure of
the organising centre and do not require any oscillatory features in the
potential.
They may be interpreted as structurally stable pre– or post–inflationary
phases with sustained scalar–geometry interactions.

\begin{figure}[t]
\centering
\tdplotsetmaincoords{70}{120}
\begin{tikzpicture}[tdplot_main_coords,scale=2.25,>=Latex,
  axis/.style={thin,->},
  orbit/.style={line width=0.8pt,postaction={decorate},
    decoration={markings,mark=at position 0.12 with {\arrow{Latex}}}},
  faint/.style={black!45,thin},
  surf/.style={black!18,thin} 
]

\draw[axis] (0,0,0) -- (2.0,0,0) node[anchor=north east] {$r$};
\draw[axis] (0,0,0) -- (0,0,1.5) node[anchor=south] {$z$};
\draw[axis] (0,0,0) -- (0,0,-1.5) node[anchor=north] {$\theta$};

\def\R{0.95}   
\def\a{0.30}   
\def\zc{0.00}  

\draw[line width=0.7pt]
  plot[domain=0:360,samples=240]
  ({(\R+\a*cos(\x))*cos(\x)},
   {\zc+(\R+\a*cos(\x))*sin(\x)},
   {\a*sin(\x)});

\draw[dashed,line width=0.6pt]
  plot[domain=0:360,samples=240]
  ({(\R-\a*cos(\x))*cos(\x)},
   {\zc+(\R-\a*cos(\x))*sin(\x)},
   {-\a*sin(\x)});

\foreach \phi in {-60,-30,0,30,60}{
  \draw[surf]
    plot[domain=0:360,samples=160]
    ({(\R+\a*cos(\phi))*cos(\x)},
     {\zc+(\R+\a*cos(\phi))*sin(\x)},
     {\a*sin(\phi)});
}

\def\k{1.5} 

\draw[black!20,dashed]
  plot[domain=0:6*pi,samples=420]
  ({(\R+\a*cos(\k*\x r))*cos(\x r)},
   {\zc+(\R+\a*cos(\k*\x r))*sin(\x r)},
   {\a*sin(\k*\x r)});

\draw[orbit]
  plot[domain=0:6*pi,samples=420]
  ({(\R+\a*cos(\k*\x r))*cos(\x r)},
   {\zc+(\R+\a*cos(\k*\x r))*sin(\x r)},
   {\a*sin(\k*\x r)});

\draw[orbit,dashed]
  plot[domain=0:360,samples=200]
  ({\R+\a*cos(\x)},  
   {0},              
   {\a*sin(\x)});    

\node[faint] at (\R+1.1*\a, 0, -2.0*\a) {\scriptsize limit cycle};

\draw[faint,->] (0.35,0.20,-0.85) arc[start angle=200,end angle=520,radius=0.18];
\node[faint] at (0.60,0.28,-0.60) {\scriptsize rotate in $\theta$};

\draw[faint,->,line width=0.6pt]
  (1.25,\zc+0.02,-0.55) -- (1.05,\zc+0.02,-0.3);
\node[faint,anchor=west] at (0.08,\zc+0.81,0.82)
  {\scriptsize invariant torus};

\end{tikzpicture}

\caption{%
Lifting of a periodic orbit to an invariant torus in the Hopf--steady--state
interaction.
A limit cycle in the reduced \((r,z)\) dynamics (dashed curve where the torus
is cut by the \((r,z)\)-plane) is rotated along the Hopf phase \(\theta\),
producing an invariant two--torus \(S^{1}\!\times\!S^{1}\) in the full
\((r,z,\theta)\) flow.
The solid winding curve illustrates a typical quasi--periodic orbit on the
torus, corresponding to persistent coupled oscillations of the Hubble mode \(z\)
and the scalar oscillation amplitude \(r\) in Case~II.}

\label{fig:torus_lift}
\end{figure}
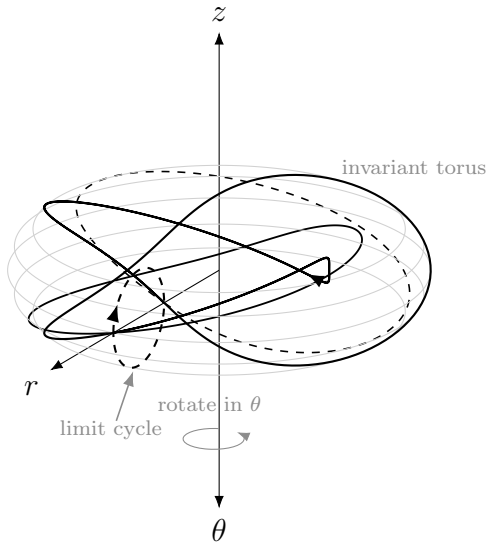

\subsection{Slow–roll observables from the unfolding}\label{subsec:slowroll}
From the definitions \eqref{eq:dimless_vars}, we work with
$y=\dot\phi/\sqrt{6}$ and $z=H$, so that for $k=0$ the
Friedmann constraint reads
\[
z^2=\tfrac13\!\left(\tfrac12\dot\phi^2+V\right)
=\tfrac13(3y^2+V).
\]
Along the slowly expanding branch of the centre
manifold we have $y=O(r)$ and $V(\phi)=V(\phi_0)+O(r^2)$, hence
$z^2=\tfrac13 V(\phi_0)+O(r^2)$.  In the scaled variables used here this gives
\begin{equation}
z^{2}=\frac{1}{3}+O(\mu_1,\mu_2,r^{2}),
\label{eq:zbackground}
\end{equation}
which is precisely the \emph{organising-centre background ansatz} at the origin:
for the bifurcation value of the parameters the equilibrium becomes
nonhyperbolic of Hopf–steady–state type.  Thus the
slowly expanding branch of the centre manifold is anchored at a de Sitter
state with $z^{2}\simeq V(\phi_{0})/3$, and at this parameter value the
equilibrium at the origin is a nonhyperbolic Hopf--steady--state point, i.e.\
the codimension-two organising centre whose versal unfolding is given by
\eqref{eq:caseI}--\eqref{eq:caseII}.

From the dynamical-systems viewpoint slow--roll (SR) and ultra slow--roll (USR)
admit a simple qualitative characterisation on the centre manifold.  In strata
such as $\chi$ the origin is a hyperbolic stable focus and the late-time
behaviour of generic initial data in its basin is SR: the background trajectory
spirals into the focus with small $(r,z)$ and the associated slow-roll
parameters satisfy $\epsilon\ll1$, $|\eta|\ll1$.  Here $\epsilon$ is the first
Hubble slow-roll parameter and $\eta$ is essentially the second Hubble
slow-roll parameter (often denoted $\epsilon_{2}$), rather than the potential
parameter $\eta_{V}\propto V''/V$.  In contrast, at the organising-centre point
$\gamma$ and along its hyperbolic continuation in the $\eta$--stratum the
origin represents a non-attractor background: it is either nonhyperbolic (at
$\gamma$) or a weakly unstable focus (in $\eta$), so trajectories that pass
close to the origin experience a finite USR episode, with $\epsilon\ll1$ but
$|\eta|=O(1)$, before peeling off towards other equilibria or oscillatory
states.  In this sense SR corresponds to a genuine hyperbolic attractor,
whereas USR appears as a transient passage near the non-attractor organising
centre and its unstable continuation.

From the perturbation point of view, this USR window corresponds to the usual
non--attractor ultra slow--roll phase, in which the comoving curvature
perturbation acquires a growing mode and the scalar power spectrum increases;
in phenomenological applications this is precisely the type of phase typically
invoked for primordial black--hole production.

The Hopf amplitude $r$ measures the small kinetic departure from de Sitter, and
the fast phase $\theta$ averages out at leading order. Since $H=z$ in our
normalisation, the exact identity for the first slow-roll parameter,
\begin{equation}
\epsilon(\tau)=-\frac{\dot H}{H^{2}},
\label{eq:epsdef}
\end{equation}
becomes, using Eq.~\eqref{eq:zbackground}, $\dot\phi^{2}=6y^{2}$, and
$y^{2}=\tfrac12 r^{2}+O(r^{3})$,
\begin{equation}
\epsilon=\frac{3\,y^{2}}{z^{2}}
     =\frac{3}{2}r^{2}+O(\mu_1,\mu_2,r^{4}),
\label{eq:epsr}
\end{equation}
so the first slow-roll parameter is directly proportional to the square of the
Hopf amplitude:
\begin{equation}
\epsilon\sim\frac{3}{2}r^{2}.
\label{eq:epsboxed}
\end{equation}
For the second slow-roll parameter,
\begin{equation}
\eta(\tau)=\frac{\dot\epsilon}{H\,\epsilon},
\label{eq:etadef}
\end{equation}
we use $r'=-\tfrac32 rz$ from the organising-centre equations. This gives
\begin{equation}
\frac{\dot\epsilon}{\epsilon}
   =\frac{d}{d\tau}(\ln r^{2})
   =-3z,
\label{eq:epsdrift}
\end{equation}
and hence
\begin{equation}
\eta = \frac{-3z}{z}
     = -3.
\label{eq:etaConstant}
\end{equation}
Therefore at the organising centre the exact Hubble parameter $\eta$ takes the
background value $\eta=-3$, originating from the geometric friction term
$3H\dot\phi$. This constant corresponds to the de Sitter decay rate and can
be absorbed into the normal--form normalisation: we consider instead the
slow residual part $\eta_{\mathrm{slow}}:=\eta+3$, whose leading contribution
is linear in the geometric mode $z$. After this normalisation and the
averaging over the fast Hopf phase already used in the derivation of
$\epsilon\simeq\tfrac32 r^{2}$, we obtain,
\begin{equation}
\eta_{\mathrm{slow}}\sim z.
\label{eq:etaboxed}
\end{equation}
To avoid extra notation we continue to denote $\eta_{\mathrm{slow}}$ simply
by $\eta$ in what follows.

To leading order, all inflationary observables depend only on the unfolding
variables $(r,z)$:
\begin{equation}
n_{s}\approx 1-6\epsilon+2\eta,
\qquad
r_{s}\approx 16\epsilon,
\qquad
A_{s}\approx \frac{H^{2}}{8\pi^{2}\epsilon},
\label{eq:inflationary}
\end{equation}
which using \eqref{eq:epsboxed}--\eqref{eq:etaboxed} become
\begin{equation}
n_{s}\approx 1-9r^{2}+2z,
\qquad
r_{s}\approx 24r^{2},
\qquad
A_{s}\approx \frac{z^{2}}{12\pi^{2}r^{2}}.
\label{eq:inflationary-rz}
\end{equation}
These formulae constitute a universal prediction of the unfolding: the observed
spectral tilt and tensor amplitude depend only on the small kinetic and
geometric modes $(r,z)$ of the organising centre, independently of any assumed
potential.

\subsubsection{Physical meaning of the unfolding parameters}

The unfolding parameters $(\mu_{1},\mu_{2})$ describe all small deformations of
the organising centre $s=1$. Physically, they correspond to the two independent
ways in which the effective inflationary dynamics can deviate from the
quadratic model: $\mu_{1}$ produces a shift/tilt deformation (breaking the
$\phi\mapsto-\phi$ symmetry and changing the relative strengths of the
      slow modes), and
$\mu_{2}$ controls the curvature and plateau behaviour of the effective
      dynamics at large field values.
In particle–physics models, these deformations typically appear as parameters
in the inflaton potential. For example, the polynomial potential \cite{klw14,kl25},
\be
V(\phi)=\frac12 m^{2}\phi^{2}(1-a\phi+b(a\phi)^{2})^{2},
\ee
has deformation parameters $(a,b)$ playing exactly the roles of
$(\mu_{1},\mu_{2})$: $a$ induces a tilt and $b$ controls the asymptotic
flattening. Our unfolding therefore provides a universal, potential–independent
classification of all such deformations. In particular, the observables
\eqref{eq:inflationary-rz} depend only on the unfolding variables $(r,z)$ and
the geometric parameters $(\mu_{1},\mu_{2})$, and not on any specific choice of
potential.

\section{Discussion}\label{sec:discussion}

The analysis developed in this paper shows that the dynamics of spatially
homogeneous scalar–field cosmology near the massless transition $s=1$ is governed
by a codimension--two organising centre of Hopf–steady--state type.  The versal
unfolding of this centre captures \emph{all} small deformations of the quadratic
model, independently of the choice of potential.  The centre manifold contains two
slow geometric modes $(r,z)$, where $r$ measures the small kinetic departure from
de Sitter and $z$ is the slow drift of the Hubble mode.  All other degrees of
freedom are either fast (the Hopf phase $\theta$) or slaved by the constraint.
This reduction is universal: any scalar potential with a nondegenerate minimum
and regular polynomial large--field behaviour induces only a reparametrisation of the
unfolding parameters $(\mu_{1},\mu_{2})$.

A key conclusion is that the inflationary observables $(n_{s},r_{s},A_{s})$ arise
directly from the dynamics of the unfolding.  The first slow–roll parameter is
proportional to the square of the Hopf amplitude, $\epsilon\simeq\tfrac32 r^{2}$,
while the second is the slow geometric mode itself, $\eta\simeq z$.  Consequently,
the spectral tilt, the tensor–to–scalar ratio, and the scalar amplitude from
Eq.~\eqref{eq:inflationary-rz} are determined solely by the unfolding variables
$(r,z)$ and the geometric parameters $(\mu_{1},\mu_{2})$, with no reference to an
underlying potential.  This identification is conceptually significant: the near
scale--invariance of primordial perturbations emerges as a \emph{geometric
property} of the organising centre, not as a feature requiring any specific
inflaton Lagrangian.

In particular, the familiar inflationary sequence of a slow--roll phase,
possible ultra slow--roll interludes, and a final oscillatory regime appears
here as a stratified bifurcation scenario.  On the centre manifold, SR
corresponds to hyperbolic strata such as $\chi$ with a unique stable focus,
USR to the organising-centre point $\gamma$ and its unstable continuation in
the $\eta$--stratum, and the oscillatory phase to the stable limit cycle on the
$\alpha$--stratum (lifting to an invariant torus in the full flow).  The Case~II
bifurcation diagram therefore encodes the allowed transitions between SR, USR
and oscillatory behaviour, while the underlying microphysics selects a
particular path in this stratified geometry. Our analysis is local in phase space and parameter space around the organising centre and does not by itself determine the global end of inflation or the total number of e–folds; these depend on how the trajectory exits the neighbourhood of the unfolding and on additional physics (e.g. reheating).

It is instructive to compare this with the potential--based approach.  In
the models of \cite{klw14,kl25}, the potential
contains two deformation parameters $(a,b)$, which control the tilt and the
large--field curvature of the potential.  These parameters play exactly the roles
of the unfolding variables $(\mu_{1},\mu_{2})$ in our setting: $a$ generates a
shift/tilt deformation breaking the $\phi\mapsto -\phi$ symmetry, and $b$
governs the deformation of the large--field behaviour away from the purely
quadratic model.  The mapping is structural: the space of potentials realising
small deformations of the quadratic model corresponds to a two--dimensional
surface in the $(\mu_{1},\mu_{2})$--plane.  Our unfolding therefore provides a
\emph{potential--independent} and \emph{canonically normalised} description of all such models.

The geometric picture that emerges is that inflation itself is a structurally
stable phenomenon associated with the unfolding of the organising centre.  The
slow variation of $(r,z)$ along the expanding branch of the centre manifold
forces $\epsilon\ll 1, \eta\ll 1$,
and hence predicts a  nearly scale--invariant spectrum of
perturbations.  In this sense, the familiar inflationary predictions arise here
as a dynamical consequence of the unfolding geometry.  If inflationary theory did
not exist, the universal behaviour encoded in the versal unfolding of the massive
scalar–field system would provide a natural mechanism for an early phase of
accelerated expansion with precisely the observed form of primordial perturbations.

This viewpoint suggests a broader interpretation: the space of inflationary
models is not a set of disparate potentials but a low--dimensional geometric
manifold parametrised by the unfolding coordinates $(\mu_{1},\mu_{2})$.  The role
of any specific potential is simply to select a path in this space.  From this
perspective the relations \eqref{eq:inflationary-rz} are universal predictions
of massive scalar–field cosmology, valid for any model lying sufficiently close to the
organising centre (i.e.\ belonging to the `versal envelope' of the massive scalar
field).  The unfolding thus provides a unifying framework for the
observational phenomenology of inflaton models, independent of their microscopic
origin.  Moreover, the same versal structure organises not only slow--roll
attractors but also ultra slow--roll transients and oscillatory torus phases, as
seen in the stratified Case~II diagram.

Finally, the existence of this organising centre and its unfolding hints at the
possibility of a more global bifurcation structure in cosmology.  As in other
areas of dynamics, the classification of neighbourhoods of singular points often
extends to more global structures, and it would be interesting to investigate
whether the scalar–field dynamics admits further degeneracies or mode
interactions beyond the Hopf–steady--state considered here.  These questions lie
at the interface of bifurcation theory, cosmology, and the theory of early--time
perturbations, and point toward a more geometric understanding of inflation.

\addcontentsline{toc}{section}{Acknowledgments}
\section*{Acknowledgments}
The research of SC was funded by RUDN University, scientific project number FSSF-2023-0003. The work of IA was supported in part by the Second Century Fund (C2F), Chulalongkorn University.

\end{document}